\documentclass[amsmath,amssymb,aps,twocolumn]{revtex4-2}

 \usepackage{amsmath, amssymb, amsfonts}
 \usepackage{mathrsfs} % בשביל \mathscr
 
\usepackage{graphicx}% Include figure files
\usepackage{dcolumn}% Align table columns on decimal point
\usepackage{bm}% bold math
\usepackage{float}
\usepackage{color}
\usepackage{braket}
\usepackage{xcolor}
\usepackage{mathtools}
\usepackage{listings}
\usepackage{float}
\usepackage[caption=false]{subfig}

\usepackage[colorlinks=true,citecolor=blue,linkcolor=blue,urlcolor=blue]{hyperref}
\usepackage{cleveref}
\usepackage{dsfont}

\newcommand{\defeq}{\vcentcolon=}
\newcommand{\norm}[1]{\left\lVert #1\right\rVert}

\begin{document}

%\preprint{APS/123-QED}

\title{Optimal Quantum Likelihood Estimation}

\author{Alon Levi}
\thanks{These two authors contributed equally.}
\author{Ziv Ossi}
\thanks{These two authors contributed equally.}
\author{Eliahu Cohen}%
\author{Amit Te'eni}%
 \email{amit.teeni@biu.ac.il}
\affiliation{Faculty of Engineering and the Institute of Nanotechnology and Advanced Materials, Bar-Ilan University, Ramat Gan 5290002, Israel
}%

\begin{abstract}
	A hybrid quantum-classical algorithm is a computational scheme in which quantum circuits are used to extract information that is then processed by a classical routine to guide subsequent quantum operations. These algorithms are especially valuable in the noisy intermediate-scale quantum (NISQ) era, where quantum resources are constrained and classical optimization plays a central role.
	
	Here, we improve the performance of a hybrid algorithm through principled, information-theoretic optimization. We focus on Quantum Likelihood Estimation (QLE) -- a hybrid algorithm designed to identify the Hamiltonian governing a quantum system by iteratively updating a weight distribution based on measurement outcomes and Bayesian inference.
	 While QLE already achieves convergence using quantum measurements and Bayesian inference, its efficiency can vary greatly depending on the choice of parameters at each step.
	
	We propose an optimization strategy that dynamically selects the initial state, measurement basis, and evolution time in each iteration to maximize the mutual information between the measurement outcome and the true Hamiltonian. This approach builds upon the information-theoretic framework recently developed in [A. Te'eni et al. Oracle problems as communication tasks and optimization of quantum algorithms, arXiv:2409.15549], and leverages mutual information as a guiding cost function for parameter selection. Our implementation employs a simulated annealing routine to minimize the conditional von Neumann entropy, thereby maximizing information gain in each iteration. The results demonstrate that our optimized version significantly reduces the number of iterations required for convergence, thus proposing a practical method for accelerating Hamiltonian learning in quantum systems. Finally, we propose a general scheme that extends our approach to solve a broader family of quantum learning problems.
\end{abstract}

\maketitle
\section{Introduction}

Hybrid quantum-classical algorithms comprise a central paradigm in near-term quantum computing, where quantum circuits are used to generate data that is processed by classical routines to guide subsequent quantum operations \cite{ge2022optimization,callison2022hybrid,nannicini2018performance,endo2020hybrid,bravyi2022hybrid}.
These algorithms are especially well-suited to the noisy intermediate-scale quantum (NISQ) era, in which the width, depth and fidelity of quantum circuits are limited, and classical post-processing is crucial for extracting meaningful results. 
While hybrid algorithms offer a promising framework for solving classically intractable problems, their performance is often highly sensitive to the quantum parameters chosen at each iteration. These parameters determine the effectiveness of the resulting measurement outcomes, which in turn determine the speed and reliability of convergence. Consequently, developing principled methods for optimizing these parameters is essential for making hybrid algorithms both practical and scalable.

In this work, we demonstrate how information-theoretic tools can be used to guide such optimization. Our goal is to extract as much information as possible from each quantum measurement. To that end, we formulate a cost function based on the mutual information between the measurement outcome and the hidden variable of interest, and use it to drive the selection of quantum circuit during the algorithm's execution.

As a concrete case study, we focus on \textit{Quantum Likelihood Estimation} (QLE)~\cite{wiebe2014hamiltonian,Wiebe2014Certification,nphys4074}, a hybrid algorithm designed for \textit{Hamiltonian learning} -- the task of identifying the unknown Hamiltonian $H$ that governs the dynamics of a quantum system \cite{granade2012robust,dutt2021active,arulandu2024survey,dutkiewicz2024advantage,renyi2024maximum}. This task plays a key role in applications such as quantum simulation, control, and device characterization~\cite{che2021learning,dutt2021active}. QLE maintains a probability distribution over a set of candidate Hamiltonians and updates this distribution iteratively using quantum measurements and Bayesian inference. Despite its simplicity and generality, QLE’s efficiency can vary greatly depending on the chosen parameters at each step~\cite{nphys4074}.

We propose an information-theoretic enhancement of QLE that systematically optimizes the choice of initial state, measurement basis, and evolution time to maximize the mutual information $I(F;Y)$ between the measurement outcome $Y$ and the unknown Hamiltonian label $F$. Our approach relies on the theoretical framework developed by Te'eni et al.~\cite{cohen2024optimization}, in which the quantum algorithm is viewed as a process of information extraction. In particular, we observe that each iteration of QLE can be interpreted as an independent oracle query problem, where the current weight vector defines the prior distribution over Hamiltonians. This interpretation enables the direct application of mutual information maximization under a single-query constraint.

To implement this idea, we define a cost function based on the conditional von Neumann entropy and employ a simulated annealing routine to minimize it over the space of quantum parameters. This leads to significant improvements in convergence speed, without compromising inference accuracy. More broadly, our work illustrates how the use of information-theoretic principles can enhance the performance of hybrid quantum algorithms in practical settings.

The remainder of this paper is structured as follows: Section~\ref{sec:Information-Theoretic-Quantities} reviews the relevant parts of the information-theoretic framework from~\cite{cohen2024optimization}. Section~\ref{sec:qle_algorithm} describes the QLE algorithm. Section~\ref{sec:optimization} details our optimization method. Section~\ref{sec:results} presents simulation results. Finally, Section~\ref{sec:conclusion} discusses the broader implications and potential extensions of this work.

\section{Information-Theoretic Tools for Algorithm Optimization}
\label{sec:Information-Theoretic-Quantities}

Hybrid quantum-classical algorithms typically proceed through multiple iterations, with each iteration involving a quantum evolution followed by a classical update. In our analysis, we isolate a single iteration and treat it as a one-query problem -- that is, the initial state undergoes a unitary transformation, passes through an oracle gate corresponding to evolution under the true Hamiltonian, and is then acted upon by a final unitary before measurement. This abstraction allows us to apply the theoretical tools developed in a previous work by some of the current authors~\cite{cohen2024optimization}, which provides an information-theoretic framework for analyzing single-query quantum algorithms. This section briefly reviews the main ingredients of this framework, which shall be used in later sections. Note that here we only discuss oracle identification problems, rather than the more general oracle \textit{classification} problems studied in \cite{cohen2024optimization}.

An \textit{oracle identification problem} consists of a set $ \mathscr{F} $ of allowed oracles, a unitary $U_f$ for each $f \in \mathscr{F}$, and a probability distribution $p_f$. A single-query algorithm prepares some initial state $ \ket{\psi_0} $; applies the quantum circuit $T_f \defeq W U_f V$ (where $V,W$ are arbitrary unitary gates); and then measure the final state $ T_f \ket{\psi_0} $ in the computational basis. The oracle identity is a random variable, denoted $F$; and the final measurement outcome is another random variable, denoted $Y$. The task is to choose $V,W$ such that the mutual information $ I \left( F;Y \right) $ is maximal. Note that one can equivalently absorb $V$ into the initial state, i.e. consider the quantum circuit $ \tilde{T}_f \defeq W U_f $ acting on $ \ket{\psi_1} $ (which would be $V \ket{ \psi_0 }$ in the original definition); this is our approach hereon.

The oracle-computer pair is modeled as a bipartite system, where the oracle is represented by a classical register keeping track of the oracle identity.
The final (pre-measurement) state of the oracle-computer system is modeled by the following classical-quantum state:
\begin{equation}\label{eq:rho_FY_full}
	\rho_{FY} = \sum_{f \in \mathscr{F}} p_f\, \ket{f} \bra{f} \otimes W U_f \ket{\psi_1} \bra{\psi_1} U_f^\dagger W^\dagger .
\end{equation}
%The resulting classical-quantum state \(\rho_{FY}\) encodes the correlations between the classical hypothesis label \(F\) and the quantum subsystem \(Y\) after one oracle query. This state serves as the starting point for our information-theoretic analysis of the learning process.
The mutual information between \(F\) and \(Y\) is given by the difference
\begin{equation}
	I(F;Y) = S(\rho_Y) - D_Y \left( \rho_{FY}; Z^{\otimes n} \right) ,
	\label{eq:mutual_information_discord_form}
\end{equation}
where \(S(\rho) = -\mathrm{Tr}[\rho \log \rho]\) is the von Neumann entropy, and \(\rho_Y = \mathrm{Tr}_F[\rho_{FY}]\) is the marginal state of subsystem \(Y\). The term $ D_Y \left( \rho_{FY}; Z^{\otimes n} \right) $ is defined as
\begin{equation}\label{eq:discord_def}
	D_Y \left( \rho_{FY}; Z^{\otimes n} \right) = S(\rho_Y) - S(\rho_{FY}) + S(\rho_{FY} \vert Z^{\otimes n}) ,
\end{equation}
where the conditional entropy \(S(\rho_{FY} \vert Z^{\otimes n})\) refers to the von Neumann entropy of the post-measurement state obtained after measuring subsystem \(Y\) in the computational basis ($n$ is the number of qubits). Note $ D_Y \left( \rho_{FY}; Z^{\otimes n} \right) $ is defined similarly to quantum discord~\cite{Bera_2018}; the only difference is that, in quantum discord, the third term of \eqref{eq:discord_def} is replaced by $ \min_\Pi S(\rho_{FY} \vert \Pi ) $, the minimum over all measurements $\Pi$. Thus, $ D_Y \left( \rho_{FY}; Z^{\otimes n} \right) $ is a basis-dependent version of quantum discord. Importantly, $ D_Y \left( \rho_{FY}; Z^{\otimes n} \right) $ is always non-negative.

In the following sections, we will demonstrate how \eqref{eq:mutual_information_discord_form} can be applied to optimize the performance of the QLE algorithm.

\section{The QLE Algorithm}
\label{sec:qle_algorithm}
The QLE algorithm is a hybrid quantum-classical algorithm for learning an unknown Hamiltonian $H_{\text{true}}$ from a set of candidate Hamiltonians $\{H_1, H_2, \dots, H_N\}$~\cite{nphys4074,dutt2021active,qi2019determining} (generally, the candidate set may be infinite; here we consider the finite case). The algorithm maintains a probability distribution over the hypotheses, encoded as a weight vector $\vec{w}^{(k)} = (w_1^{(k)}, w_2^{(k)}, \dots, w_N^{(k)})$, which is initialized uniformly (or according to prior knowledge) and updated throughout the execution.

At each iteration, the quantum circuit prepares the system in a fixed state $\ket{\psi_1}$, typically the computational basis state $\ket{0}$. The system is then evolved under the unitary operator
$ U(t) = e^{-i H_{\text{true}} t} $.
The time parameter $t$ is selected dynamically using the \textit{Particle Guess Heuristic} (PGH)
$ t = \frac{1}{ \norm{ H_i -H_j}_2 } $,
where $H_i, H_j$ are two Hamiltonians sampled according to the current weights~\cite{wiebe2014hamiltonian,Wiebe2014Certification}. Intuitively, this heuristic increases discrimination between Hamiltonians by selecting a time that amplifies their dynamical differences.
Following time evolution, the state is rotated using a fixed unitary matrix $W$, and a projective measurement is performed in the computational basis, yielding a classical outcome $Y \in \{0, 1\}^n$ (for a Hamiltonian acting on $n$ qubits).

Then, each candidate Hamiltonian $H_j$ is used to simulate the corresponding expected outcome probabilities:
\begin{equation}
	p_j^{(a)} = \left| \bra{a} W e^{-i H_j t} \ket{\psi_1} \right|^2, \quad a \in \{0, 1\}^n.
\end{equation}
Letting $Y \in \left\{ 0,1 \right\}^n$ denote the observed measurement result, the likelihood of observing $Y$ under hypothesis $H_j$ is given by $ \mathcal{L}_j = p_j^{(Y)} $.
The weights are then updated using Bayesian inference:
\begin{equation}
	w_j^{(k+1)} = \frac{w_j^{(k)} \cdot \mathcal{L}_j}{\sum_{i=1}^{N} w_i^{(k)} \cdot \mathcal{L}_i},
\end{equation}
ensuring that the distribution remains normalized at each iteration.

The algorithm proceeds until one of the weights approaches unity, indicating high confidence in the corresponding Hamiltonian hypothesis. Note that two Hamiltonians differing by a constant (i.e. $H_2 = H_1 + c \mathds{1}$, where $ c \in \mathbb{R} $ and $\mathds{1}$ is the identity operator) cannot be discerned: they generate the exact same dynamics (up to an unphysical global phase), hence the same probabilities $p_j^{(a)}$.

\section{Optimizing QLE using our framework}
\label{sec:optimization}

In this section we describe an optimized version of the original QLE algorithm described in the previous section.
Consider the problem of learning a Hamiltonian from a restricted set $\{H_1, \dots, H_N\}$ ($N$ is the number of hypotheses under consideration). Each Hamiltonian $H_f$ in this ensemble is a $2 \times 2$ Hermitian matrix acting on a single qubit.
The algorithm's performance is measured as $ \sum_{f=1}^{N} \# \mathrm{iterations} \left( f \right) $; here $ \# \mathrm{iterations} \left( f \right) $ denotes the number of iterations the algorithm takes to succeed, in a run where the true Hamiltonian is $H_{\text{true}} = H_f$. We define success as the condition in which the weight associated with the correct Hamiltonian exceeds $0.99$, indicating high confidence in the correct identification. If the algorithm converges to the \textit{wrong} Hamiltonian, we define $ \# \mathrm{iterations} \left( f \right) = \infty $.

At the beginning of each run, the weight distribution is initialized uniformly, with $w_j^{(0)} = 1/N$ for all $j = 1, \dots, N$.
The initial state is a general single-qubit state:
\begin{equation}
	\ket{\psi_1} = \cos(\alpha) \ket{0} + e^{i\beta} \sin(\alpha) \ket{1},
\end{equation}
where $\alpha \in [0, \pi]$ and $\beta \in [0, 2\pi]$ are free parameters that define the orientation of the qubit on the Bloch sphere. $ \ket{\psi_1} $ then evolves under the true Hamiltonian, i.e. we apply $U(t) = e^{-i H_{\text{true}} t}$, where $t \in [0, 2\pi / \Delta_{\min} ]$ is the evolution time parameter and $ \Delta_{\min} $ is the minimal energy gap in the given set $\{H_1, \dots, H_N\}$.
The post-query unitary $W$ is given by:
\begin{equation}
	W = \begin{bmatrix}
		\cos(\theta/2) & e^{-i\phi} \sin(\theta/2) \\
		e^{i\phi} \sin(\theta/2) & -\cos(\theta/2) 
	\end{bmatrix} ,
\end{equation}
with $\theta \in [0, \pi]$ and $\phi \in [0, 2\pi]$ controlling the measurement basis rotation. %The quantum evolution is applied via the operator $U(t) = e^{-i H_{\text{true}} t}$, where $t \in [0, 2\pi]$ is the evolution time parameter.

A central component of our methodology is the dynamic selection of the five parameters $(\alpha, \beta, \theta, \phi, t)$ at each iteration. This is performed using a dedicated optimization routine called \texttt{QLE\_optimization}, which is described in detail in the following section. Besides the selection of these parameter, our optimized version of QLE proceeds precisely as the original.

%For each Hamiltonian $H_i$ in the set, we repeat this simulation process and track the evolution of the weight vector. We define success as the condition in which the weight associated with the correct Hamiltonian exceeds $0.99$, indicating high confidence in the correct identification.

The key insight that underlies \texttt{QLE\_optimization}, is that each iteration of the QLE algorithm can be regarded as an independent \textit{single-query oracle problem} of the type discussed in \Cref{sec:Information-Theoretic-Quantities}. In this formulation, the unknown oracle $ f \in \mathscr{F} $ corresponds to the true Hamiltonian, while the prior distribution over hypotheses $p_f$ is represented by the current weight vector. 
This oracle-based interpretation allows us to frame each iteration as a single-query information extraction task, where the objective is to maximize the mutual information between the measurement outcome and the unknown Hamiltonian. Crucially, this formulation enables the direct application of the results of \Cref{sec:Information-Theoretic-Quantities}. Note that~\cite{granade2012robust} introduced a similar framework for Hamiltonian learning using  optimal control settings; however, they only considered the evolution time $t$ as a controlled parameter. More recently, \cite{dutt2021active} proposed optimizing control settings to reduce query complexity, including the initial state and measurement basis. A key difference between their work and ours is in our measure for algorithms' success. We define success as identifying the true Hamiltonian with sufficiently high fidelity; on the other hand, \cite{dutt2021active} aims to minimize the distance (in parameter space) between the guess and the true Hamiltonian. These two notions of success yield two distinct cost functions that are optimized in each query: mutual information in our work, contrasted with Fisher information in \cite{dutt2021active}.

Now, we can describe how to implement the function \texttt{QLE\_OPTIMIZATION}:
Our goal is to choose the parameters $(\alpha, \beta, \theta, \phi, t)$ such that the mutual information $I(F;Y)$ between the true Hamiltonian $F$ and the observed measurement outcome $Y$ is maximized. Intuitively, maximizing this quantity ensures that each measurement outcome carries as much information as possible about the correct hypothesis, thereby accelerating convergence.

We begin by noting that \eqref{eq:rho_FY_full} is a spectral decomposition of $\rho_{FY}$, since the states $ \ket{f} $ are orthogonal. Thus, the probabilities $p_f$ are the eigenvalues of $ \rho_{FY} $, hence $ S(\rho_{FY}) = H \left( F \right) $. Since the probabilities $p_f$ in a given iteration are independent of the parameters $\left( \alpha, \beta, \theta, \phi, t \right)$ (for that same iteration), we treat them as constants; hence $H \left( F \right)$ is also considered constant. Substituting $H \left( F \right) = \mathrm{const.} $ into \eqref{eq:discord_def} and then substituting that into \eqref{eq:mutual_information_discord_form}, we obtain:
\begin{equation}
	I(F;Y) = \mathrm{const.} - S \left( \rho_{FY} \vert Z \right).
\end{equation}
Thus, we observe that maximizing mutual information is equivalent to minimizing the conditional von Neumann entropy $ S \left( \rho_{FY} \vert Z \right) $. This is the cost function used in our optimization routine.

To optimize the parameters that minimize $ S \left( \rho_{FY} \vert Z \right) $, we employ a simulated annealing algorithm \cite{kirkpatrick1983optimization,cerny1985thermodynamical,aarts1988simulated,geman1984stochastic,ingber1993simulated}.
The optimization begins with a randomly initialized parameter vector $ \bm{x} = [t, \theta, \phi, \alpha, \beta] $, where $ t $ is the evolution time, $ \theta $ and $ \phi $ define the measurement basis, and $ \alpha, \beta $ specify the initial quantum state. At each iteration, the algorithm generates a set of new candidate vectors -- called ``neighbors'' -- by adding random perturbations scaled by the current ``temperature'' $ T $ to each parameter:
\begin{equation}
	\bm{x}_{i+1} = \bm{x}_i + T \cdot \mathrm{rand}([-1,1]^5) \circ (\mathrm{range}/2) ,
	\label{eq:annealing_update}
\end{equation}
where \(\circ\) denotes the Hadamard (element-wise) product, and \(\mathrm{range}/2\) is a vector containing half the range of each parameter. This ensures that the random perturbations scale appropriately for each parameter.
The parameters are wrapped into their valid ranges:
\begin{equation*}
	\theta, \alpha \in [0, \pi), \quad t, \phi, \beta \in [0, 2\pi)
\end{equation*}
using modular arithmetic.

The cost function is evaluated for each candidate neighbor. The candidate with the lowest cost is chosen and accepted if its cost is lower than the current point’s  (i.e., $ \mathrm{cost}_{\text{new}} < \mathrm{cost}_{\text{old}} $). Otherwise, a candidate with a higher cost may still be accepted with probability
\begin{equation}
	\exp\left( \frac{\mathrm{cost}_{\text{old}} - \mathrm{cost}_{\text{new}}}{T} \right),
\end{equation}
where $T$ is the current ``temperature''. This probabilistic acceptance criterion allows the algorithm to escape local minima during early stages of the annealing process.

The ``temperature'' is then reduced according to a geometric cooling schedule:
\begin{equation}
	T \leftarrow \alpha \cdot T, \quad \alpha \in (0, 1),
\end{equation}
gradually focusing the search on high-quality regions of the parameter space. The parameter $\alpha$ controls the rate of cooling: values close to 1 lead to slower cooling, allowing broader exploration and help escaping local minima;, while lower values cause the algorithm to converge more quickly. To balance exploration and convergence, $\alpha$ is typically chosen in the range $[0.90, 0.97]$. A slower cooling schedule (i.e., higher $\alpha$) increases the likelihood of escaping local minima and reaching a global optimum. The value of $\alpha$ we used is 0.9.
The process is repeated for a fixed number of iterations, and the best parameter set found is selected.

While simulated annealing effectively minimizes the entropy in our simulations, other optimization strategies may also be applicable. For example, exhaustive grid search with fine resolution over all five parameters yields slightly suboptimal but still strong results, with the advantage of simplicity and reproducibility at the cost of computation time.
 
\section{Simulation results}
\label{sec:results}

To evaluate the effectiveness of our information-guided optimization approach, we compared it against a baseline version of the QLE algorithm in which no adaptive parameter optimization is performed. In the baseline version, the parameters $ (\alpha, \beta, \theta, \phi) $ are scanned exhaustively over a discrete multi-dimensional grid. This corresponds to a nested-loop search that evaluates all possible combinations, enabling a brute-force exploration of the parameter space.
The evolution time $ t $ is selected using the standard PGH heuristic. The same configuration is used for all iterations and all Hamiltonians.
%For each Hamiltonian in the set $ \left\{H_1, \dots, H_4\right\} $, we recorded the minimum number of iterations required to achieve convergence.
We begin with a simple, yet non-trivial choice of the candidate Hamiltonians -- Pauli operators and scaled versions thereof: $ H_1 = \sigma_x $, $ H_2 = 2\sigma_x $, $ H_3 = \sigma_z $, and $ H_4 = 2\sigma_z $. %This family of Hamiltonians was chosen to produce a simple -- yet nontrivial -- instance of a Hamiltonian learning problem (here ``nontrivial'' means the Hamiltonian cannot be learned in a single query with probability $1$).
We compared the total number of iterations required for all four Hamiltonians to converge, $ \sum_{f=1}^4 \# \mathrm{iterations} \left( f \right) $.

\begin{figure}
	\centering
	% Subfigure (a)
	\subfloat[]{
		\centering
		\includegraphics[width=\linewidth]{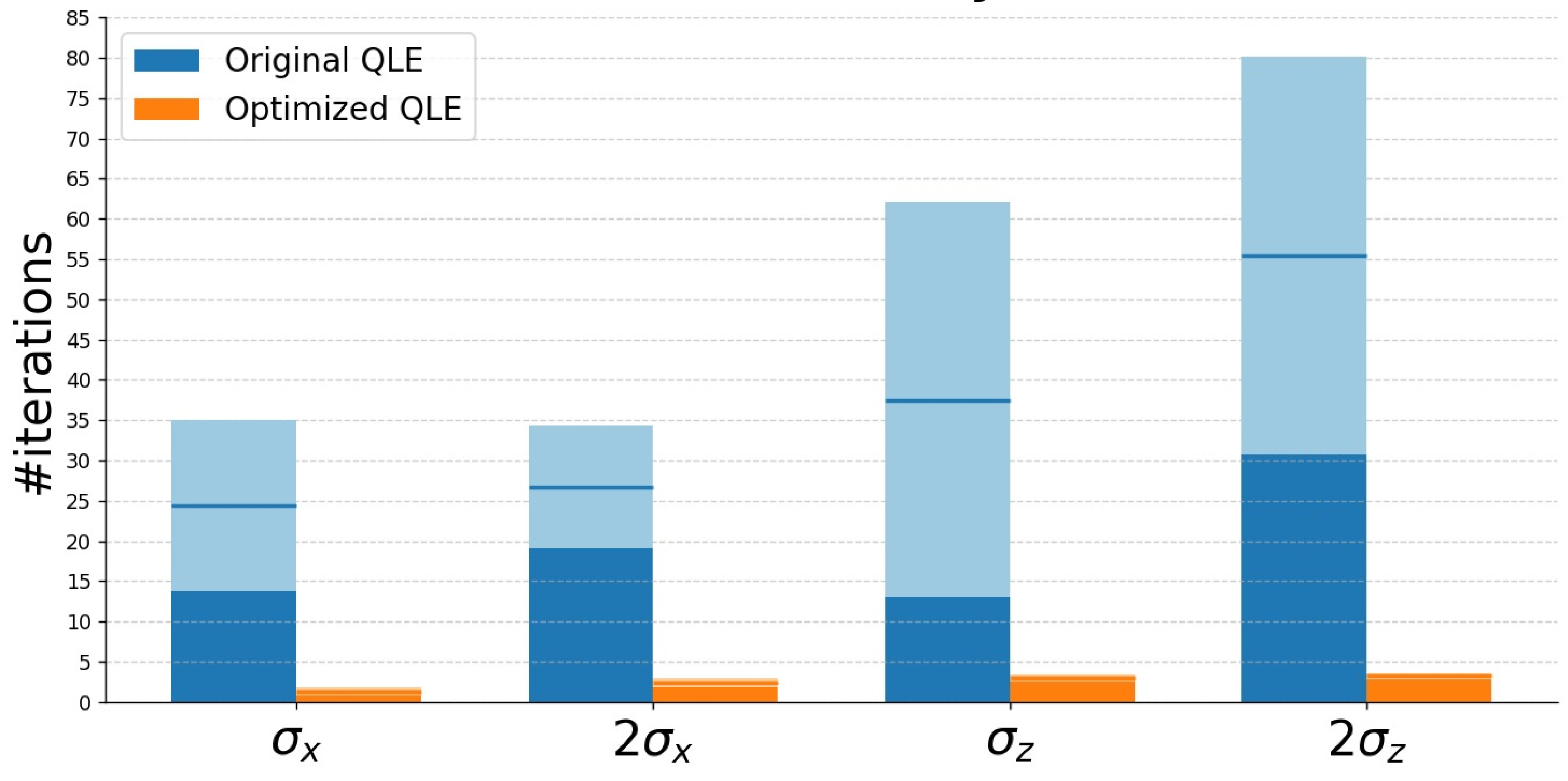}
		%\caption{}
		\label{fig:fig_a}
	}
	
	\vspace{1em} % vertical spacing between the two subfigures
	
	% Subfigure (b)
	\subfloat[]{
		\centering
		\includegraphics[width=\linewidth]{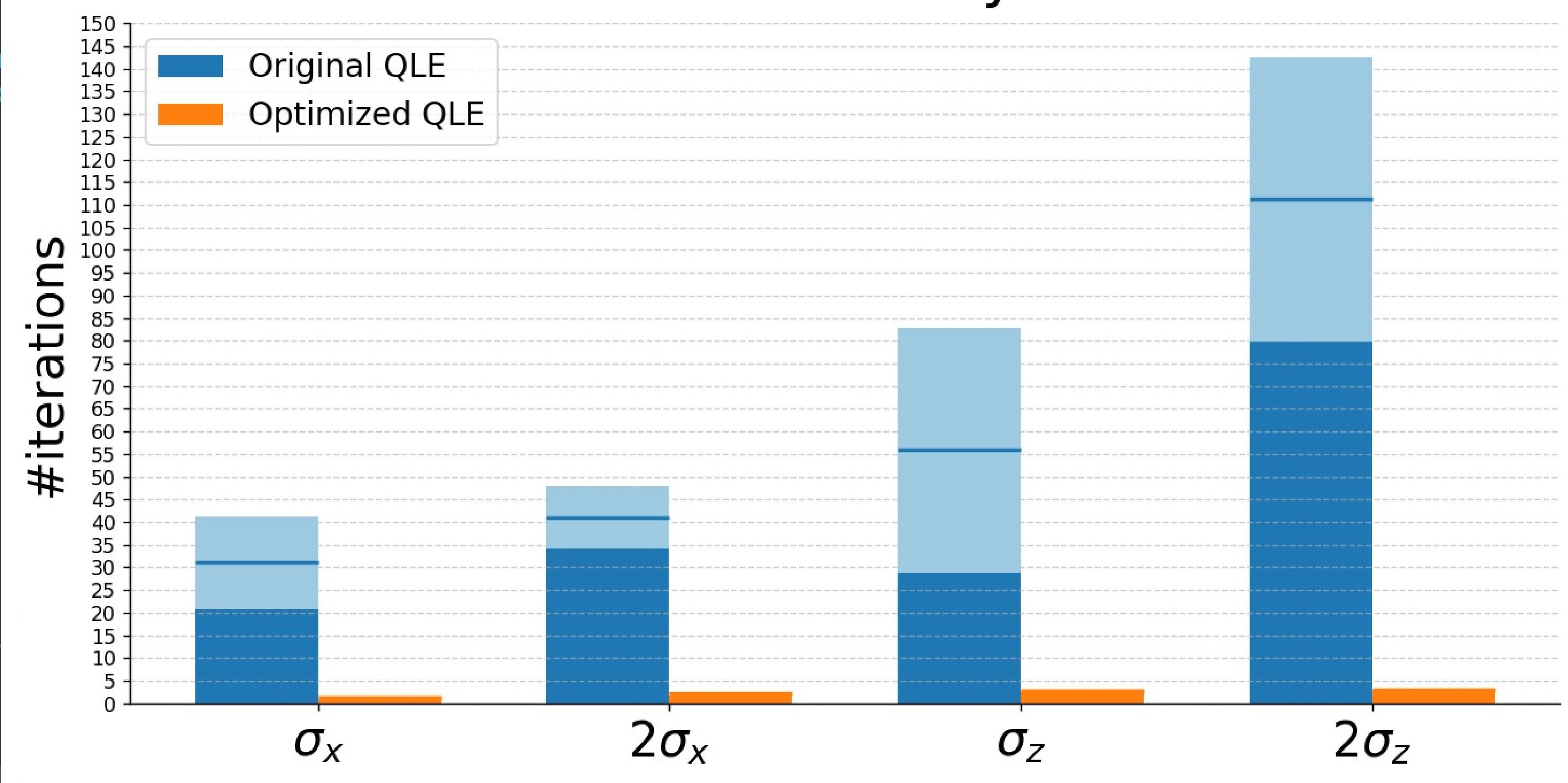}
		%\caption{}
		\label{fig:simulated_annealing_suls}
	}
	
	\caption{
		Comparison between the performance of the original and optimized (simulated-annealing-version) QLE algorithms.
		The horizontal axis represents the four Hamiltonians.
		Each Hamiltonian is shown with two adjacent bars: the left bar corresponds to the original QLE and the right bar to the optimized QLE.
		The darker central segment of each bar indicates the mean number of iterations, while the lighter shaded region above and below it represents the range of one standard deviation (\(\pm \text{Std. Dev.}\)) from the mean.
		Panel (a) shows the case with the threshold being 0.99, while panel (b) shows the stricter threshold requiring the maximum weight to exceed 0.9999.
		The simulation code is available at~\cite{MyCodeRepo}.
	}
	\label{fig:qle_comparison}
\end{figure}

This analysis revealed that the best static configuration required on average $144$ iterations for all four Hamiltonians to converge. In contrast, our adaptive version of QLE -- which dynamically selects the parameters at each iteration using simulated annealing -- required only $9$ iterations to achieve convergence. This represents a substantial reduction in the number of oracle queries and highlights the practical advantage of our approach. The number of iterations required by each method for every Hamiltonian is presented in \Cref{fig:fig_a}.
Moreover, when we required the maximum weight to exceed $0.9999$ instead of $0.99$, the advantage of our optimization method became even more pronounced, as shown in \Cref{fig:simulated_annealing_suls}.

To isolate the specific contribution of the optimization strategy itself, we also evaluated a variant of our approach. In this version, the parameters were optimized using the same discrete grid search employed in the baseline QLE implementation. 
However, unlike the original version where the parameters were fixed throughout the run, here the grid search was repeated at every iteration -- thereby adapting the parameters continuously. This eliminates the advantage our method might gain from simply having access to a wider parameter space (since simulated annealing yields parameters unconfined to any grid), allowing for a more direct comparison that focuses solely on the effectiveness of the optimization technique used.
In this approach, convergence was achieved within 10 iterations. This suggests that the primary factor driving the efficiency of the learning process is our optimization strategy, and not the enlarged parameter space. 

We also considered the following set of six Hamiltonians:
\begin{equation*}
	\left\{
	\begin{aligned}
		&\sigma_x + \sigma_z, \quad \mathrm{diag}(1.2,-0.8), \quad
		2\sigma_y + \mathrm{diag}(1,2),\\
		&\frac{4}{\sqrt{2}}
		\begin{bmatrix}
			1 & 1 \\
			1 & -1
		\end{bmatrix}, \quad
		\sigma_x + \frac{4}{\sqrt{2}}
		\begin{bmatrix}
			1 & 1 \\
			1 & -1
		\end{bmatrix}, \quad
		\begin{bmatrix}
			1 & 2 \\
			2 & -1
		\end{bmatrix}
	\end{aligned}
	\right\} .
\end{equation*}
For this set, the original QLE failed to converge for all cases, indicating that several of these Hamiltonians are indistinguishable under its fixed measurement strategy. In contrast, the optimized QLE converged correctly for all six, requiring in average only $4-5$ iterations per Hamiltonian. Unlike the aforementioned four-Hamiltonian set $\{\sigma_x,\,2\sigma_x,\,\sigma_z,\,2\sigma_z\}$, which the original QLE could solve due to clear separability in measurement outcomes, this larger set includes mixed-Pauli combinations and asymmetric shifts that obscure distinguishing features. The optimized method overcomes these limitations, achieving both reliable identification and markedly higher efficiency.

\section{Discussion}
\label{sec:conclusion}
In this work, we showed how hybrid quantum algorithms can be optimized using tools from information theory. By leveraging quantities such as mutual information and conditional entropy, we formulated a principled approach to dynamically selecting algorithmic parameters in order to extract maximal information from each quantum measurement.

As a case study, we applied this methodology to the QLE algorithm, demonstrating that optimizing the initial state, measurement basis, and evolution time based on information-theoretic criteria significantly accelerates convergence. We illustrated this in the setting of single-qubit Hamiltonian learning, but the approach extends naturally to multi-qubit systems. In that regime, however, the number of parameters grows exponentially with the number of qubits. This ``curse of dimensionality'' could potentially be mitigated by a fully quantum variant of our method, in which the optimization subroutine is carried out using quantum annealing~\cite{Ray1989Sherrington,Apolloni1989Quantum,apolloni1990numerical,FINNILA1994343,Kadowaki1998Quantum,Farhi2001Quantum,Rajak2023Quantum} rather than classical simulated annealing. Although quantum annealing has been shown to outperform simulated annealing for certain optimization problems~\cite{Santoro_2006,morita2008mathematical,Heim2015Quantum,yan2022analytical}, further investigation is needed to determine whether it yields an advantage in our setting.

While our examples focused on Hamiltonian learning tasks with a finite candidate set, the same methodology can be extended to continuous families of Hamiltonians ${H_x}$, indexed by parameters $x \in [a,b]$. For a desired precision $\epsilon$, one can discretize the parameter range into bins of length $\epsilon$, thereby reducing the problem to classifying the unknown Hamiltonian $H_x$ according to its bin. Each iteration can still be regarded as a single-query oracle problem, but now the task falls into the broader class of oracle \textit{classification} problems rather than the oracle identification problems discussed in \Cref{sec:Information-Theoretic-Quantities}. This perspective allows us to apply the tools of~\cite{cohen2024optimization} in a more general setting, leading to a similar optimization framework.

Importantly, QLE serves here as a simplified setting designed to illustrate, in principle, how the information-theoretic framework introduced in~\cite{cohen2024optimization} can be used to improve the performance of hybrid quantum-classical algorithms. The same methodology can naturally be extended to more general settings in which measurement outcomes guide iterative decision-making.
In the general case, the learning problem comprises the following building blocks:
\begin{itemize}
	\item An initial state $ \rho_i = \rho_i \left( \bm{c}_i, \bm{h}_i \right) $;
	\item A completely positive, trace-preserving (CPTP) map $ \mathcal{P} = \mathcal{P} \left( \bm{c}_p, \bm{h}_p \right) $;
	\item The final state $ \mathcal{P} \left\{ \rho_i \right\} $;
	\item An \textit{attribute} of the final state, $ a \left( \mathcal{P} \left\{ \rho_i \right\}, \bm{c}_a, \bm{h}_a \right) $.
\end{itemize}
The initial state, CPTP map and attribute all depend on two types of parameters: \textit{controlled} and \textit{hidden} parameters, which are labeled $ \bm{c} $ and $\bm{h}$ respectively. The attribute is a map from the final state and parameters to a probability distribution. Since the final state is a function of the initial state and CPTP map, the attribute is a function of all parameters, $ a = a \left( \bm{c}_i, \bm{h}_i, \bm{c}_p, \bm{h}_p, \bm{c}_a, \bm{h}_a \right) $.
The hidden parameters' values are unknown, but are drawn from a known probability distribution; and the task is to compute some function of the hidden parameters, $ f \left( \bm{h}_i, \bm{h}_p, \bm{h}_a \right) $, using as few queries as possible. In each query we specify the values of the controlled parameters $ \bm{c}_i, \bm{c}_p, \bm{c}_a $ and sample the probability distribution given by the final state's attribute.

This problem strictly generalizes the Hamiltonian learning setup outlined in \Cref{sec:qle_algorithm}. Therefore, we propose an algorithm that generalizes our approach for optimizing QLE. First, we maintain a ``weight'' (probability distribution) for every hidden parameter, reflecting our current knowledge. The weights are initialized as the known prior distribution, and updated after every query based on the gained information, via Bayesian or likelihood-based inference. In each query we choose the controlled parameters $ \bm{c}_i, \bm{c}_p, \bm{c}_a $ such that the following mutual information is maximized:
\begin{equation*}
	I \left( a \left( \bm{c}_i, \bm{h}_i, \bm{c}_p, \bm{h}_p, \bm{c}_a, \bm{h}_a \right) ; f \left( \bm{h}_i, \bm{h}_p, \bm{h}_a \right) \right) .
\end{equation*}
For any fixed weights of the hidden parameters, the mutual information is only a function of the controlled parameters; and we can try finding its maximum via simulated annealing.
This adaptive ``prepare-measure-update-optimize'' loop ensures that every query yields maximal utility for refining the parameter estimates. As such, this approach offers a principled route to accelerating convergence in a wide class of quantum learning and estimation tasks beyond the QLE example considered here. Noisy scenarios can also be seamlessly accommodated within this framework. Any noise model affecting state preparation, evolution, or measurement can be captured by augmenting $\bm{h}_i$, $\bm{h}_p$ or $\bm{h}_a$ with additional hidden parameters. The learning task can be defined either to include these noise parameters or to focus solely on the original ones.

\begin{acknowledgments}
	We are grateful to Pawe\l{} Horodecki, Adi Makmal, Marcin Nowakowski, Maria Schuld and Zohar Schwartzman-Nowik for helpful discussions.
	This work was partially supported by the European Union's Horizon Europe research and innovation programme under grant agreement No. 101178170 and by the Israel Science Foundation under grant agreement No. 2208/24.
\end{acknowledgments}

%\bibliography{ArticleReview}% Produces the bibliography via BibTeX.

%\appendix
%\renewcommand\thesubsection{\thesection.\arabic{subsection}}
%\onecolumngrid

\end{document}